\documentclass[a4paper,11pt]{article}
\usepackage{pos}

\title{CEPC-on-Gaussino: an application of Gaussino simulation framework for CEPC experiment}

\author*[a]{Tao Lin}
\author[a,b]{Weidong Li}
\author[c]{Xingtao Huang}
\author[c]{Teng Li}
\author[a]{Ziyan Deng}
\author[a]{Chengdong Fu}
\author[a]{Jiaheng Zou}

\affiliation[a]{Institute of High Energy Physics, Chinese Academy of Sciences,\\
  19B Yuquan Road, Shijingshan District, Beijing, China}
\affiliation[b]{University of Chinese Academy of Sciences, 19A Yuquan Road, Shijingshan District, Beijing, China}
\affiliation[c]{Institute of Frontier and Interdisciplinary Science, Shandong University, Qingdao, Shandong, China}

\emailAdd{lintao@ihep.ac.cn}

\abstract{The Circular Electron Positron Collider (CEPC) is a future Higgs factory to measure the Higgs boson properties. Like the other future experiments, the simulation software plays a crucial role in CEPC for detector designs, algorithm optimization and physics studies. Due to similar requirements, the software stack from the Key4hep project has been adopted by CEPC. As the initial application of Key4hep, a simulation framework has been developed for CEPC based on DD4hep, EDM4hep and k4FWCore since 2020. However, the current simulation framework for CEPC lacks support for the parallel computing. To benefit from the multi-threading techniques, the Gaussino project from the LHCb experiment has been chosen as the next simulation framework in Key4hep. This contribution presents the application of Gaussino for CEPC. The development of the CEPC-on-Gaussino prototype will be shown and the simulation of a tracker detector will be demonstrated.}

\FullConference{42nd International Conference on High Energy Physics (ICHEP2024)\\
18-24 July 2024\\
Prague, Czech Republic\\}

\begin{document}
\maketitle

\section{Introduction}
The Circular Electron Positron Collider (CEPC) is an $\mathrm{e^+}$$\mathrm{e^-}$ Higgs factory designed to produce Higgs, W, Z and top quarks, with the goal of discovering new physics beyond the Standard Model. In November 2018, the CEPC Conceptual Design Report (CDR) was released \cite{CEPCStudyGroup:2018rmc, CEPCStudyGroup:2018ghi}. Following this, the CEPC Accelerator Technical Design Report (TDR) was published in in December 2023 \cite{CEPCStudyGroup:2023quu}. The CEPC team is currently working on the Reference Detector TDR, which is scheduled for released in June 2025. To support the detector design and physics performance studies, the CEPC software (CEPCSW) plays a crucial role by providing simulation and reconstruction tools to the end users. 

The CEPCSW is developed based on a common turnkey software stack called Key4hep \cite{Key4hep:2021zms}. The primary goal of Key4hep is to maximize the sharing of software components among different experiments. Currently, Key4hep is utilized by CEPC, CLIC \cite{Aicheler:2012bya, Linssen:2012hp}, FCC \cite{FCC:2018byv, FCC:2018evy, FCC:2018vvp}, ILC \cite{Behnke:2013xla, ILC:2013jhg} and other future experiments. By adopting the Key4hep software stack, CEPCSW is composed of three main parts: the underlying external libraries, the core software and the CEPC-specific applications. 

The core software provides the key functionalites to the applications. The Gaudi framework \cite{Barrand:2001ny} defines the interfaces for all software components and control their execution. EDM4hep \cite{Gaede:2022leb} defines a generic event data model, while K4FWCore manages the event data. DD4hep \cite{Frank:2014zya} provides the geometry description. In addition to these software components, there are CEPC-specific software elements, such as the simulation framework. This framework includes the physics generator interface, Geant4 simulation and beam background mixing. 

\begin{figure}[h]
	\centering
	\includegraphics[width=.8\textwidth]{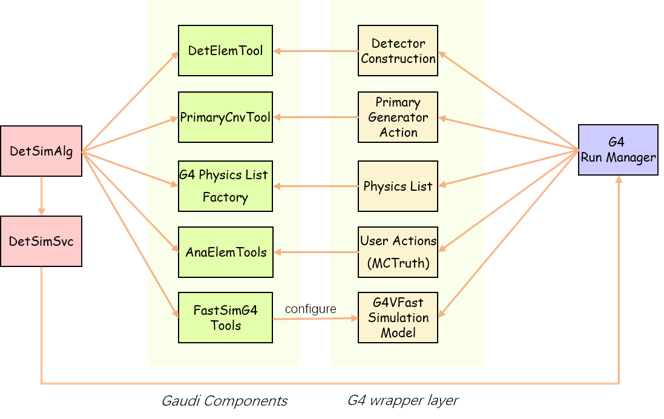}
	\caption{Design of simulation framework}
	\label{fig:current-simulation-design}
\end{figure}

Figure \ref{fig:current-simulation-design} illustrates the design of the simulation framework. The entry point is a Gaudi algorithm called \texttt{DetSimAlg}, which comprises a list of Gaudi tools responsible for detector construction, primary particle generation, physics list, user actions and fast simulation. Corresponding Geant4-derived classes are implemented to integrate the Gaudi tools with Geant4. These Geant4-derived classes are registered with the Geant4 Run Manager during the initialization stage. The detector construction tool and physics list tool are invoked at beginning of the event loop. During the event loop, \texttt{DetSimAlg} calls the Geant4 Run Manager to simulate an event, which in turn invokes the appropriate primary generator action tools and user action tools. 

\section{Moving to a new simulation framework}

\begin{figure}[h]
	\centering
	\includegraphics[width=.8\textwidth]{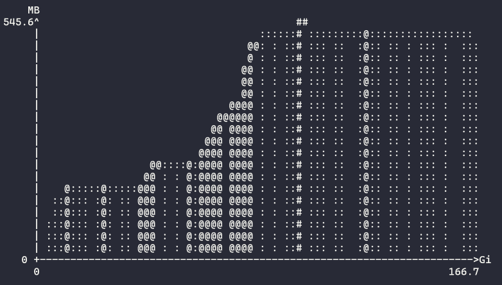}
	\caption{Heap memory usage of simulation in CEPCSW}
	\label{fig:heap-memory}
\end{figure}

The current simulation framework does not support multi-threading. Multi-threading simulation could significantly reduce the memory usage by sharing geometries and the physics list. The memory usage of the current simulation was measured using the CEPC detector option \texttt{TDR\_o1\_v01} and the Geant4 physics list \texttt{QGSP\_BERT}. To measure the memory usage at initialization, 100 single muon events were simulated to minimize memory usage during the event loop. The resident set size (RSS) memory is approximately 950 MB at the initialization stage. Figure \ref{fig:heap-memory} shows the heap memory usage, measured with the profiler tool Valgrind Massif. According to the Massif report, most of the memory is used by geometry related components, such as the ROOT \texttt{TGeo} \cite{Brun:1997pa} geometries used by DD4hep, the Geant4 geometry converted by \texttt{DDG4}, and Geant4 geometry voxelization during \texttt{G4GeometryManager::BuildOptimisations}. Therefore, implementing multi-threaded simulation could reduce memory usage by sharing these components. 

\begin{figure}[h]
	\centering
	\includegraphics[width=.8\textwidth]{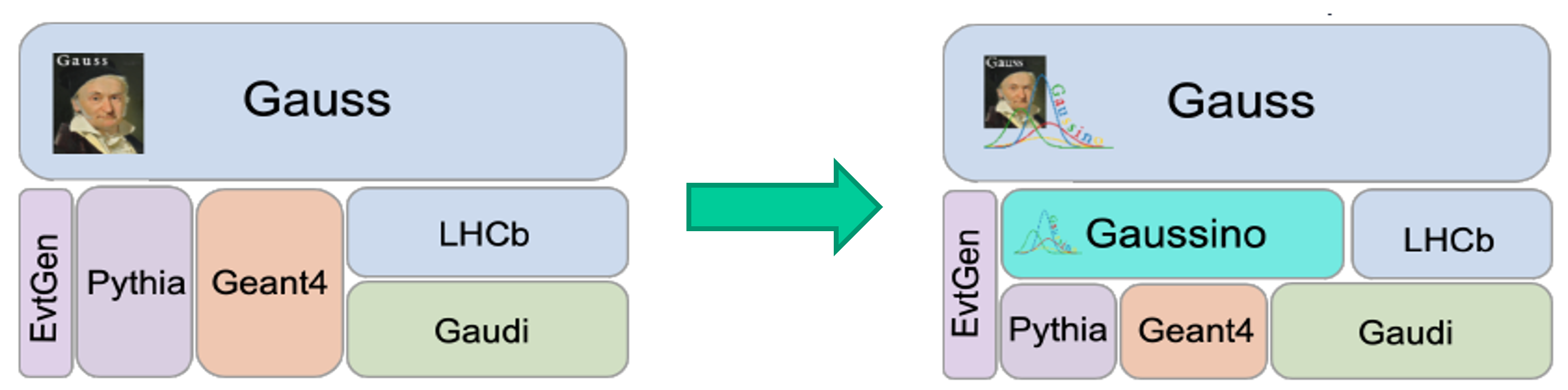}
	\caption{Evolution of LHCb simulation framework}
	\label{fig:gauss-on-gaussino}
\end{figure}

There are two options for a multi-threaded simulation framework. One option is to develop a new framework based on the current serial version. The other option is to adopt an existing framework. Gaussino \cite{Mazurek:2022tlu} is a potential solution for simulation within Key4hep project \cite{Key4hep:2023xhe}. Gaussino is a simulation framework from LHCb. As shown in Figure \ref{fig:gauss-on-gaussino}, the LHCb simulation framework Gauss is divided into the common parts, named Gaussino, and the LHCb dedicated parts, named the Gauss-on-Gaussino. The underlying framework is based Gaudi Functional and Gaudi Hive, which provide better support for multi-threading. Therefore, Gaussino has been chosen as the next simulation framework in CEPCSW, and the CEPC-on-Gaussino prototype has been proposed.

\section{CEPC-on-Gaussino prototype}
\begin{figure}[h]
	\centering
	\includegraphics[width=.6\textwidth]{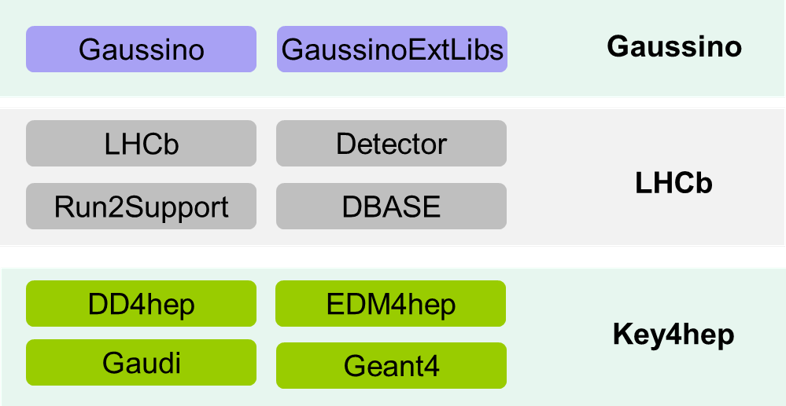}
	\caption{The dependencies of Gaussino}
	\label{fig:gaussino-stacks}
\end{figure}

One of the challenges in adopting Gaussino in CEPCSW is the existing dependencies on LHCb software, and the decoupling of these dependencies is still on going. Figure \ref{fig:gaussino-stacks} shows the current software dependencies. To develop the CEPC-on-Gaussino prototype, there are three steps: 
\begin{enumerate}
    \item Using Gaussino with the full LHCb software;
    \item Creating a modified version with fewer LHCb dependency; 
    \item Directly using the Key4hep version without LHCb dependency, which is not available at the moment.
\end{enumerate}

The first step involes using Gaussino with LHCb software dependencies. Since the external libraries are deployed at \texttt{/cvmfs}, building Gaussino with LHCb software installation script is relatively straightforward. During this step, only the geometries from CEPCSW are used in Gaussino, while all the underlying software is based on LHCb software. To load the CEPC tracker detector, \texttt{DD4hepCnvSvc} is adopted as the geometry service instead of the LHCb detector description library. As there is no detector response implemented, the Geant4 stepping verbose information are printed to verify whether the simulation works. Multi-threading is also tested with multiple cores.

The second step involves creating a modified version with fewer LHCb dependencies. During this step, all the underlying software is based on CEPC software. Gaussino and its dependencies are built as external libraries of CEPC software. Special branches in the repositories are created, and the \texttt{CMakeLists.txt} files are modified to build Gaussino. Only the necessary packages from LHCb software are built, such as the Event Data Model \texttt{GenEvent} and \texttt{MCEvent}. After optimization, approximately 20 shared libraries are built in total. 

\begin{figure}[h]
	\centering
	\includegraphics[width=.9\textwidth]{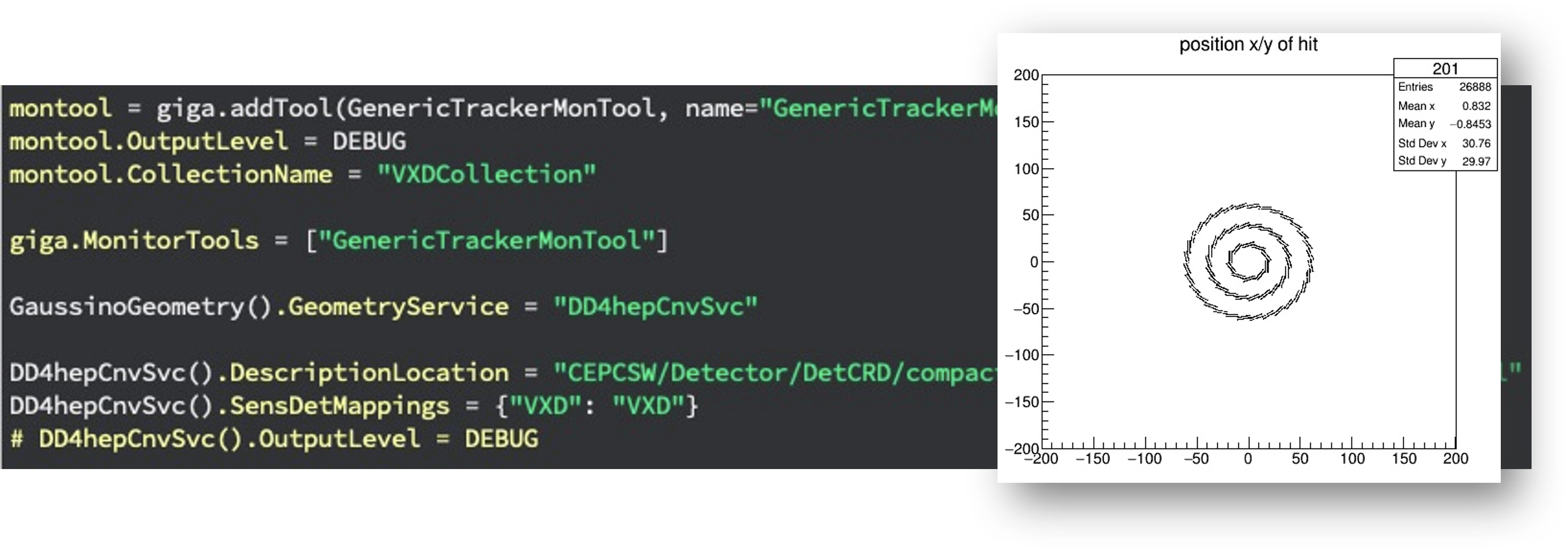}
	\caption{The monitoring tool in job options and the histogram of positions}
	\label{fig:monitoring}
\end{figure}

The detector responses are implemented for the tracker detector. Classes from \texttt{DDG4} are reused with minor modifications, such as the hit class \texttt{Geant4Hit} and the sensitive detector base class \texttt{DDG4SensitiveDetector}. A concrete sensitive detector class \texttt{GenericTrackerSensitiveDetector} is then implemented. To integrate the sensitive detector into Gaussino, an additional Gaudi factory named \texttt{GenericTrackerSensDetTool} is used to create the object. A monitoring tool is also implemented to save the user output, including histograms of positions and deposit energies. In the monitoring tool, an instance of \texttt{G4Event} is provided at the end of each event, the hit collections are retrieved and converted to the \texttt{Geant4Hit}. Then these hit information are filled into the histograms. Firgue \ref{fig:monitoring} shows the job options and the corresponding histogram. 

\section{Summary}
The simulation framework in CEPCSW has been developed to support the CEPC Reference Detecotor TDR. To benefit from multi-threading techniques and reduce memory usage, it is necessary to transition from a serial version to a multi-threaded version. The Gaussino project from LHCb has been chosen as the simulation framework within Key4hep. Consequently, the CEPC-on-Gaussino prototype has been proposed and implemented without dependencies on the whole LHCb software stack.

\acknowledgments
This work is supported by the NSFC Basic Science Center Program for ``Joint Research on High
Energy Frontier Particle Physics'' (Grant No. 12188102), the Youth Innovation Promotion Association of CAS (No. 2022011), and the European Union's Horizon 2020 Research and Innovation programme under Grant Agreement No 101004761.


\bibliographystyle{JHEP}
\bibliography{skeleton}

\end{document}